# Vector solitons in nonlinear lattices

Yaroslav V. Kartashov,[1] Boris A. Malomed,[2] Victor A. Vysloukh,[3] and Lluis Torner[1]

[1]*ICFO-Institut de Ciencies Fotoniques, and Universitat Politecnica de Catalunya, Mediterranean Technology Park, 08860 Castelldefels (Barcelona), Spain*

[2]*Department of Physical Electronics, School of Electrical Engineering, Faculty of Engineering, Tel Aviv University, Tel Aviv, 69978, Israel*

[3]*Departamento de Fisica y Matematicas, Universidad de las Americas – Puebla, Santa Catarina Martir, 72820, Puebla, Mexico*

We consider two-component solitons in a medium with a periodic modulation of the nonlinear coefficient. The modulation enables the existence of complex multihump vector states. In particular, vector solitons composed of dipole and fundamental, or dipole and even double-hump components exist and may be stable. Families of unstable scalar solitons can be stabilized in the vectorial form, due to the coupling to a stable second component.

*OCIS codes: 190.0190, 190.6135*

Self-sustained nonlinear modes in the form of solitons appear in a variety of materials and settings. Transverse modulation of parameters of the nonlinear material substantially affects properties of solitons. The modulation restricts the soliton's mobility, but it may also lead to appearance of new families of solutions that do not exist in uniform media. Recently, solitons in materials with periodic modulation of the nonlinearity have drawn attention. Such purely nonlinear lattices are interesting because their effective strength depends on the energy of the nonlinear excitation. Properties of single-component solitons in harmonic nonlinear lattices (NLs) have been analyzed in detail [1-5], and more sophisticated nonlinearity landscapes were explored too [6-8]. Solitons may also form in materials where linear and nonlinear lattices coexist and compete with each other [9-14]. The competition results in power-controlled transformations of the soliton's shape, modification of stability, and enhancement of the mobility.

To date, studies of solitons in NLs have focused on scalar excitations. However, solitons may include several components (optical beams with different polarizations or carrier wavelengths, or different atomic states in Bose-Einstein condensates, BECs) that may bind into vector states via the cross-phase-modulation. The vectorial coupling greatly enriches soliton families and alters their stability [15-20]. In NLs, only the simplest two-component solutions combining odd or even modes have been considered [21]. Here we demonstrate that the sinusoidal NL supports vector solitons composed of dipole and fundamental, or dipole and even components, which are stable in a large part of their existence domains. Families of unstable scalar solitons may be stabilized in the vectorial form, due to the coupling to a stable second component.

The evolution of two coupled wave packets in the cubic medium with the NL obeys the following equations for field amplitudes $q_1$ and $q_2$:

$$i\frac{\partial q_{1,2}}{\partial \xi} = -\frac{1}{2}\frac{\partial^2 q_{1,2}}{\partial \eta^2} + \sigma(\eta)q_{1,2}(|q_{1,2}|^2 + C|q_{2,1}|^2). \tag{1}$$

Here $\xi$ is the distance (or time in BEC), and $\eta$ is the transverse coordinate; the nonlinearity coefficient is $\sigma = \sigma_{\mathrm{m}}\cos^2(\Omega\eta) - 1$, where $\Omega$ is the modulation frequency, $C$ is the cross-modulation coefficient (in optics, $C = 1$ for mutually incoherent light beams, $C = 2$ for beams with different carrier wavelengths, and $C = 2/3$ for two orthogonal linear polarizations). Equation (1) conserves the total and partial energy flows: $U = U_1 + U_2 \equiv \int_{-\infty}^{\infty}(|q_1|^2 + |q_2|^2)d\eta$, which also determine the energy sharing: $S_{1,2} = U_{1,2}/U$. We choose scaling $\Omega = 2$, mostly concentrating on the case of $\sigma_{\mathrm{m}} = 1$.

Nonlinear lattices might be realized in optics as large-period photonic crystals whose holes are filled with an index-matching liquid [22], or in photorefractive materials where the periodically varying nonlinearity may be created by indiffusion of dopants [23]. Notice that in an externally biased photorefractive medium with a periodic background illumination with intensity $I_{\mathrm{bg}}(\eta)$, the nonlinear contribution to the refractive index is proportional to $E_{\mathrm{sc}} \approx E_0[1 - I/I_{\mathrm{bg}}(\eta)]$ for $I \ll I_{\mathrm{bg}}$ (here $I$ is the intensity of the probe beam, and $E_0$ is the biasing field), which also leads to a nonlinearity modulation.

We search for solutions to Eq. (1) as $q_{1,2}(\eta,\xi) = w_{1,2}(\eta)\exp(ib_{1,2}\xi)$ with propagation constants $b_{1,2}$. The stability was analyzed for solutions $q_{1,2} = [w_{1,2} + (u_{1,2} + iv_{1,2})\exp(\delta\xi)]\exp(ib_{1,2}\xi)$, where $u_{1,2}$ and $v_{1,2}$ are small perturbations with growth rate $\delta = \delta_r + i\delta_i$. After the substitution of this into Eq. (1) and linearization, one arrives at an eigenvalue problem, which can be solved numerically:

$$\delta u_{1,2} = -\frac{1}{2}\frac{d^2 v_{1,2}}{d\eta^2} + [\sigma(w_{1,2}^2 + Cw_{2,1}^2) + b_{1,2}]v_{1,2},$$
$$\delta v_{1,2} = +\frac{1}{2}\frac{d^2 u_{1,2}}{d\eta^2} - [\sigma(3w_{1,2}^2 + Cw_{2,1}^2) + b_{1,2}]u_{1,2} - 2\sigma C w_1 w_2 u_{2,1},$$
(2)

First we remind basic properties of scalar solitons ($q_1 \neq 0$, $q_2 = 0$) in NLs. Fundamental, even, and dipole scalar solitons are depicted in Fig. 1(a). At low values of the energy flow the fundamental and even solitons expand across the entire lattice, getting localized with the increase of $U$. Importantly, solitons in NLs do not bifurcate from Bloch waves (NL does not affect the beam at $U \to 0$), therefore they do not exhibit oscillations as $U \to 0$ [Fig. 1(b)] in contrast to solitons in linear lattices. Dipole modes exist above a threshold value of $U$. Scalar fundamental solitons are always stable, even ones are always unstable, while dipole modes become stable above a certain critical value of $b$.

Coupled fundamental and dipole modes form asymmetric vector solitons [Fig. 2(a)]. To explore them, we fix $b_2$ and vary $b_1$. The fundamental-dipole solution becomes strongly asymmetric with the decrease of propagation constant $b_1$ of the dipole component: The left lobe of $w_1(\eta)$ gradually vanishes, while its right lobe shifts to the right. Increasing $b_1$ leads to a gradual equilibration of the lobes in the $w_1$ component, while $w_2$ develops two almost identical humps. As a result, the fundamental-dipole vector soliton transforms into a bound state of the even-dipole type. Thus, at fixed $b_2$, the fundamental-dipole soliton complex exists for $b_1^{\text{low}} \leq b_1 \leq b_1^{\text{upp}}$. At $C=1$ the energy flow of the dipole component is always lower than that of the fundamental one [Fig. 2(b)], but $S_1$ approaches $S_2$ as $b_1 \to b_1^{\text{upp}}$, until the soliton jumps onto the even-dipole branch. The energy flow grows with $b_1$ everywhere, except for a narrow region close to $b_1^{\text{low}}$, with $dU/db_1 < 0$ [Fig. 2(c)]. The existence domain of the fundamental-dipole solitons expands almost linearly with $b_2$ [Fig. 2(d)], but such solutions cannot be found below a critical value of $b_2$. Interestingly, at $C=1$, the fundamental-dipole solitons are stable almost in the entire domain of their existence, except for the above-mentioned narrow region where $dU/db_1 < 0$. However at $C<1$ an extended instability domain appears near $b_1^{\text{low}}$, while at $C>1$ solutions are unstable near both cutoffs. The stability domain gradually shrinks as $C$ grows, making the vector soliton unstable already at $C=1.4$. Notice that variation in $\sigma_{\text{m}}$ does not notably affect properties of fundamental-dipole solitons: For given $b_1$, $b_2$, and $C=1$, the energy flow slowly grows with increase of $\sigma_{\text{m}}$; at fixed $b_2$ the existence domain in terms of $b_1$ expands with $\sigma_{\text{m}}$, while solutions can be found in both cases $\sigma_{\text{m}} \to 0$ and $\sigma_{\text{m}} \gg 1$.

The second type of vector solitons involves coupled dipole and even components [Fig. 3(a)]. At $C=1$ the existence domain of such soliton is rather narrow. With the increase of $b_1$, the dipole component becomes stronger, while its even counterpart vanishes, and at $b_1 = b_1^{\mathrm{upp}}$ the vector soliton transforms into the scalar dipole. The even component becomes more pronounced if $b_1$ decreases, and one gets an even scalar soliton at $b_1 = b_1^{\mathrm{low}}$ [Fig. 3(b)]. Such scenario takes place for $C \leq 1.02$, while for $C > 1.02$ the picture is different, with the $w_1$ component vanishing with the increase of $b_1$. Cutoff $b_1^{\mathrm{low}}$ grows linearly with $b_2$, but solutions are not found below a critical value of $b_2$ [Fig. 3(c)]. Remarkably, the cross-modulation coupling with the stable dipole component may result in *stabilization* of the even component, which is unstable as a scalar. At $C=1$ the stabilization takes place for $b_1$ close to $b_1^{\mathrm{upp}}$. Figure 3(d) shows the stability domain in the $(b_1, b_2)$ plane. Solitons exist in the narrow region to the left of the red curve. Their existence domain shrinks with the increase of $b_2$, but the even-dipole solitons always remain stable in a considerable part of their existence region, which is adjacent to the upper cutoff. The fundamental-dipole solitons considered above transform into even-dipole solitons exactly at the value of $b_1$ where the even-dipole solitons become completely stable. The minimal width of the existence domain for even-dipole solitons is achieved around $C = 1.02$ [Fig. 3(e)]. Since at $C > 1.02$ the dipole component vanishes at $b_1 \to b_1^{\mathrm{upp}}$, the stabilization of the vector soliton occurs near the lower cutoff, $b_1^{\mathrm{low}}$, in contrast to the case of $C \leq 1.02$, where the stabilization takes place near the upper cutoff. Figure 3(f) shows a dependence of the real part of the perturbation growth rate on $b_1$. Decreasing $\sigma_{\mathrm{m}}$ causes a significant expansion of the existence domain for even-dipole solitons at $C = 1$. When $\sigma_{\mathrm{m}}$ is too small and NL is weak, the two humps in the even component become less pronounced and may fuse into a single hump when $b_1 \to b_1^{\mathrm{low}}$. The stability domain remains narrow with decreasing $\sigma_{\mathrm{m}}$. For $C = 1$ and $b_2 = 3$, the existence domain of the even-dipole solitons ends at $\sigma_{\mathrm{m}} \approx 1.7$.

Summarizing, we have considered families of two-component solitons supported by NLs with the cubic nonlinearity. The existence and stability regions are found for the vector solitons built of the dipole mode in one component, and a fundamental or even (double-hump) one in the other. The interaction with the stable dipole component may stabilize the even mode, which is always unstable in the scalar case.

# References with titles

# References without titles

# Figure captions

Figure 1. (a) Profiles of fundamental, dipole, and even scalar solitons with $b_1 = 1.5$. In gray regions $\sigma(\eta) < -1/2$, white in white ones $\sigma(\eta) > -1/2$. (b) Energy flow versus $b_1$ for the fundamental and even solitons. Circles correspond to the solitons shown in (a).

Figure 2. (a) The profile of the fundamental-dipole vector soliton at $b_1 = 1.9$, $b_2 = 3$. (b) Energy sharing between the components versus $b_1$ at $b_2 = 3$. (c) Energy flow versus $b_1$. The circle corresponds to the soliton shown in (a). (d) The existence domain in the $(b_2, b_1)$ plane. In this and next figures $C = 1$, except for 3(d).

Figure 3. (a) The profile of the even-dipole vector soliton at $b_1 = 2.67$, $b_2 = 3$. (b) The energy sharing versus $b_1$ at $b_2 = 3$. (c) Lower cutoff $b_1^{\text{low}}$ for the even-dipole soliton versus $b_2$. Stability (white) and instability (shaded) domains in the $(b_1, b_2)$ plane (d), and in the $(C, b_1)$ plane at $b_2 = 3$ (e). (f) $\delta_r$ versus $b_1$ at $b_2 = 3$.

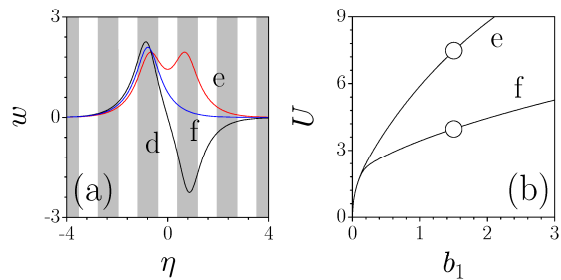

Figure 1. (a) Profiles of fundamental, dipole, and even scalar solitons with $b_1 = 1.5$. In gray regions $\sigma(\eta) < -1/2$, white in white ones $\sigma(\eta) > -1/2$. (b) Energy flow versus $b_1$ for the fundamental and even solitons. Circles correspond to the solitons shown in (a).

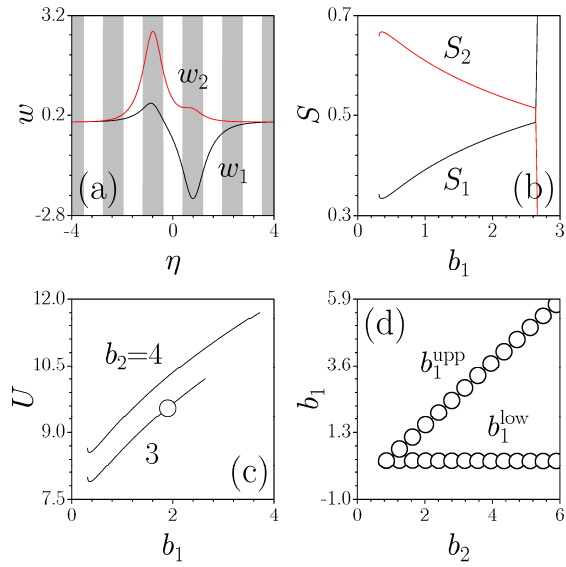

Figure 2.   (a) The profile of the fundamental-dipole vector soliton at $b_1 = 1.9$, $b_2 = 3$. (b) Energy sharing between the components versus $b_1$ at $b_2 = 3$. (c) Energy flow versus $b_1$. The circle corresponds to the soliton shown in (a). (d) The existence domain in the $(b_2, b_1)$ plane. In this and next figures $C = 1$, except for 3(d).

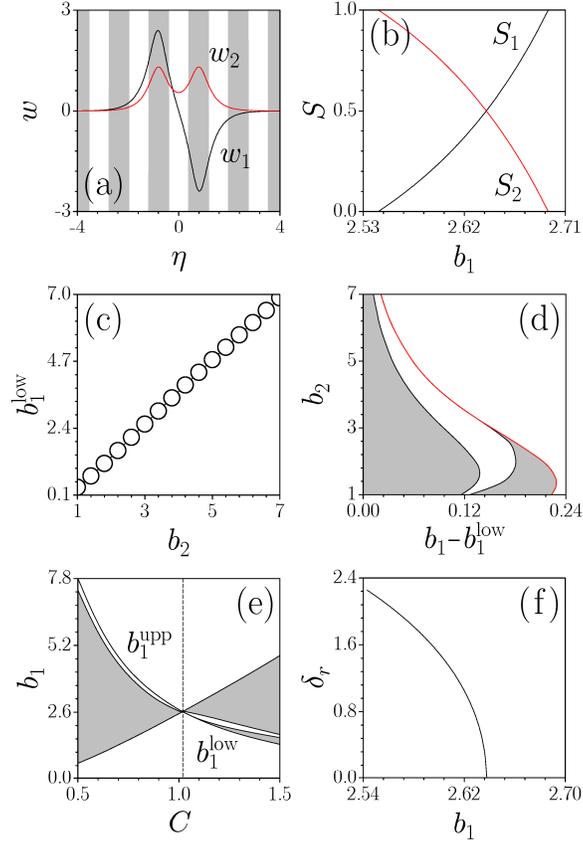

Figure 3. (a) The profile of the even-dipole vector soliton at $b_1 = 2.67$, $b_2 = 3$. (b) The energy sharing versus $b_1$ at $b_2 = 3$. (c) Lower cutoff $b_1^{\text{low}}$ for the even-dipole soliton versus $b_2$. Stability (white) and instability (shaded) domains in the $(b_1, b_2)$ plane (d), and in the $(C, b_1)$ plane at $b_2 = 3$ (e). (f) $\delta_r$ versus $b_1$ at $b_2 = 3$.